\documentclass[sigconf]{acmart}
\usepackage[utf8x]{inputenc}
\usepackage{setspace,enumitem}
\setlist{leftmargin=5mm}
\usepackage{amsmath, amssymb, latexsym, siunitx, nicefrac, textcomp, color}
\usepackage{graphicx, epstopdf}
\usepackage{IEEEtrantools}

\usepackage[capitalise]{cleveref}
\usepackage{booktabs}
\usepackage{tabularx}
\newcolumntype{Y}{>{\centering\arraybackslash}X}
\newcolumntype{R}{>{\raggedleft\arraybackslash}X}
\newcolumntype{L}{>{\raggedright\arraybackslash}X}
\usepackage{multirow, threeparttable}
\usepackage{ragged2e}% \Centering
\usepackage{makecell}% \thread

\usepackage{hyphenat}
\usepackage{algorithm}
\usepackage{algpseudocode}

\bstctlcite{IEEE:ForceEtal}

\begin{document}
\setcopyright{none}
\acmDOI{}
\acmISBN{}
\acmConference[WOSET'18]{Workshop on Open-Source EDA Technology}{November 8, 2018}{San Diego, CA, United States}
\acmYear{}
\copyrightyear{}
\acmArticle{}
\acmPrice{}
\pagestyle{plain} % Removes running headers

\title{DATC RDF: An Open Design Flow from Logic Synthesis to Detailed Routing}

\author{Jinwook Jung}
\affiliation{%
  \institution{Korea Advanced Institute of Science and Technology}
  \city{Daejeon}
  \country{Korea}
}
\email{jinwookjung@kaist.ac.kr}

\author{Iris~Hui-Ru~Jiang}
\affiliation{%
  \institution{National Taiwan University}
  \city{Taipei}
  \country{Taiwan}
}
\email{huiru.jiang@gmail.com}

\author{Jianli Chen}
\affiliation{%
  \institution{Fuzhou University}
  \city{Fuzhou}
  \country{China}
}
\email{jlchen@fzu.edu.cn}

\author{Shih-Ting Lin}
\affiliation{%
  \institution{National Chiao Tung University}
  \city{Hsinchu}
  \country{Taiwan}
}
\email{cxzasd3661512@gmail.com}

\author{Yih-Lang Li}
\affiliation{%
  \institution{National Chiao Tung University}
  \city{Hsinchu}
  \country{Taiwan}
}
\email{ylli@cs.nctu.edu.tw}

\author{Victor N.\ Kravets}
\affiliation{%
  \institution{IBM T.\ J.\ Watson Research Center}
  \city{New York}
  \country{United States}
}
\email{kravets@us.ibm.com}

\author{Gi-Joon Nam}
\affiliation{%
  \institution{IBM T.\ J.\ Watson Research Center}
  \city{New York}
  \country{United States}
}
\email{gnam@us.ibm.com}

\keywords{VLSI design flow, CAD contest, physical design}

\begin{abstract}
In this paper, we present DATC Robust Design Flow (RDF) from logic synthesis to detailed routing.
Our goals are 1) to provide an open-source academic design flow from logic synthesis to detailed routing based on existing contest results, 2) to construct a database for design benchmarks and point tool libraries, and 3) to interact with industrial designs by using industrial standard design input/output formats.
We also demonstrate RDF in a scalable cloud infrastructure.
Design methodology and cross-stage optimization research can be conducted via RDF.
\end{abstract}

\maketitle

%------------------------------------------------------------------------------
% Introduction
%------------------------------------------------------------------------------
\section{Introduction}\label{sec:intro}
EDA research contests and their released benchmark suites have successfully attracted research endeavors on timely and practical problems.
These contests stimulated innovative solutions which indeed advanced the cutting edge technologies.

Based on these outstanding point tools, the \textit{DATC Robust Design Flow (RDF)} is developed to provide an open-source academic design flow, which can facilitate design methodology and cross-stage optimization research.
Our goals are 1) to provide an academic reference flow from logic synthesis to detailed routing based on existing contest results, 2) to construct a database for design benchmarks and point tool libraries, and 3) to interact with industrial designs by using industrial standard design input/output formats.

%------------------------------------------------------------------------------
% Overall Flow
%------------------------------------------------------------------------------
\section{DATC Robust Design Flow}\label{sec:overall_flow}
\begin{figure}[ht!]%
        \centering
        \includegraphics[width=0.95\linewidth]{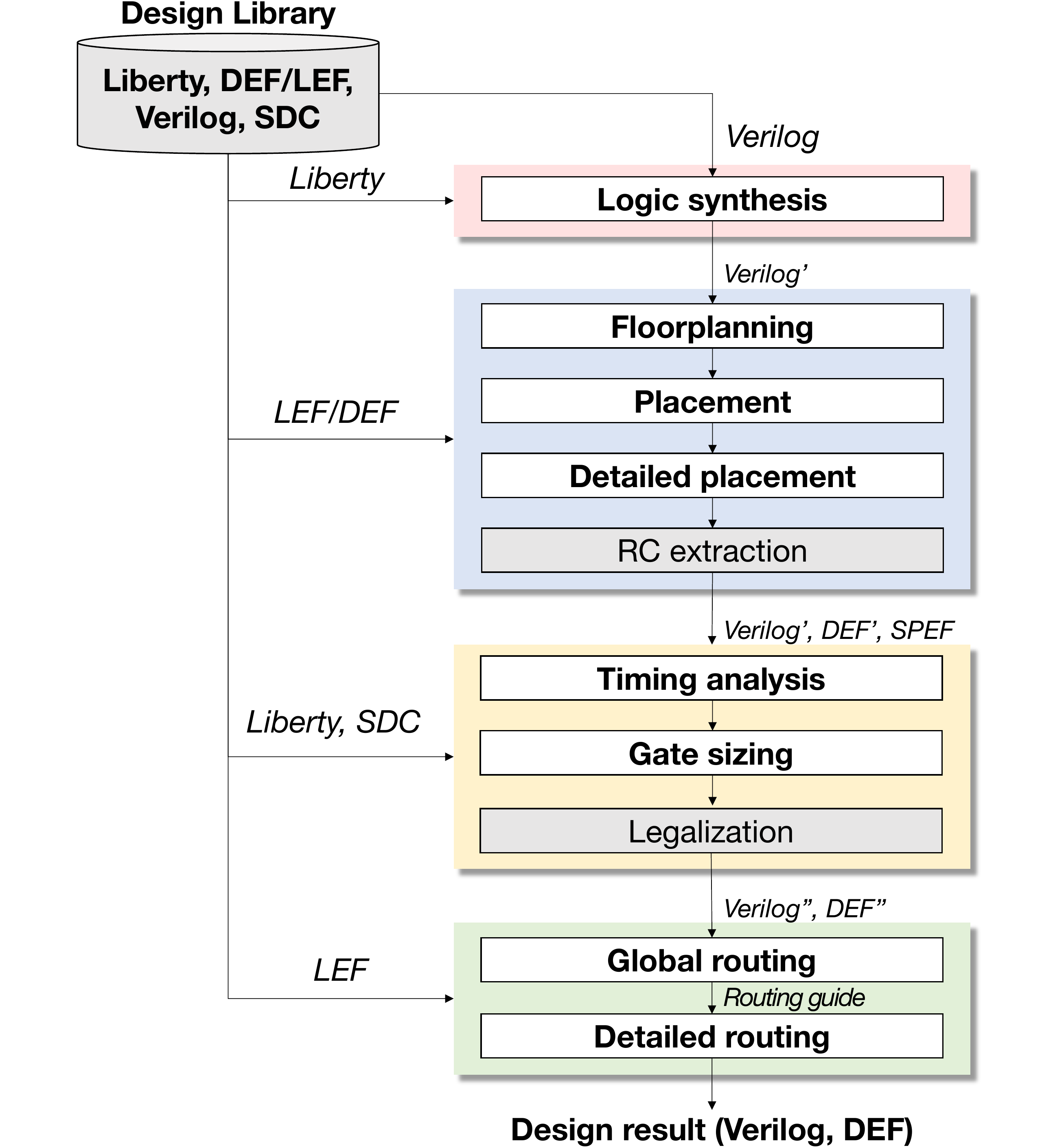}%
        \caption{Overview of DATC Robust Design Flow.}%
        \label{fig:overall_flow}%
    \end{figure}

DATC RDF improves the preliminary versions~\cite{jung:2016od,jung:2017dr} to deliver a complete research infrastructure of VLSI design flow~\cite{datc_rdf_repo}.
Our goal is to provide an open-source academic design flow that covers entire design stages, i.e., from logic synthesis to detailed routing, based on the public academic point tools from the previous EDA research contests~\cite{Nam:2005gx,Nam:2008gr,Ozdal:2012gs,Kim:2014td,tau2017,Darav:2017,Mantik:2018is},
\Cref{fig:overall_flow} illustrates the overview of DATC RDF.
It includes academic point tools for logic synthesis, global placement, detailed placement, timing analysis, gate sizing, global routing, and detailed routing.
These tools are interfaced via transition scripts that enable data exchange between tools of other domains.

A design library for DATC RDF contains:
\begin{itemize}[noitemsep]
    \item A circuit written in a structural \textbf{Verilog} netlist.
    \item Standard cell library in \textbf{Liberty} format.
    \item Physical information of standard cells along with technology information in \textbf{LEF} format, which defines physical dimensions of each cell.
    \item Initial floorplan described in \textbf{DEF} format.
    \item Design constraints in \textbf{SDC} (Synopsys Design Constraints) format, such as clock period, driver information of each input port, and load capacitance of each output.
\end{itemize}

Given a design library, DATC RDF starts with the logic synthesis and generates a logic-optimized gate-mapped Verilog netlist.
Taking the netlist and LEF/DEF from the design library, global and detail placements are then performed.
Wire parasitics are extracted so that the timing of the placement result can be analyzed.
Gate sizing may optionally run to remove timing violations while minimizing leakage power.
Legalization is called to remove illegal placement caused by the gate sizing.
Finally, global routing and detailed routing is performed.

Currently, RDF database contains:
\begin{enumerate}[noitemsep]
    \item Benchmarks: 2017 TAU Contest, and IWLS 2005 Benchmarks.
    \item Logic synthesis: ABC.
    \item Global placers: NTUPlace3, ComPLx, mPL5/6, Capo, FastPlace3-GP, Eh?Placer.
    \item Detailed placers: FastPlace3-DP, MCHL.
    \item Global routers: NCTUgr, FastRoute4.1, BFG-R.
    \item Detailed routers: NCTUdr.
    \item Gate sizers: USizer 2013 and USizer 2012.
    \item Timers: OpenTimer, iTimerC2.0.
    \item Cell libraries: ISPD 2012/2013 Contests, ASAP 7nm library.
\end{enumerate}
Users can customize their own flow based on the above options.

%------------------------------------------------------------------------------
% Updates in this version
%------------------------------------------------------------------------------
\section{Updates In This Version}
In this section, we highlight the extension from the preliminary versions~\cite{jung:2016od,jung:2017dr} of DATC RDF as follows.
Details of the logic synthesis, global placement, gate sizing and global routing stages can be found in~\cite{jung:2016od}.

\subsection{Technology and Standard Cell Libraries}
To fully cover the entire VLSI design flow, technology and standard cell libraries have to contain cell timing library (for logic synthesis, gate sizing, and timing analysis), as well as technology information and design rules (for placement and routing).
In this regard, two technology libraries are available in the current implementation of RDF.
The default technology library is a variant of ISPD 2012/2013 Discrete Gate Sizing Contests~\cite{Ozdal:2012gs,Ozdal:2013ea}.
We bring the Liberty standard cell library from the contest library.
Since the ISPD 2012/2013 contest benchmark suite does not include a LEF file, which includes technology information and physical dimension of each cell, we take the LEF file generated by the A2A methodology presented in~\cite{Kahng:2014fk}.

Another library that DATC RDF supports is the ASAP 7nm library~\cite{Vashishtha:2017ap,Xu:2017sc}.
We take a total of 89 standard cells from the library, which includes basic combinational gates along with some complex gates such as AOI221 or OAI222.
Only a basic D-type flip-flop with no reset and set ports is included in our library because the logic synthesis tool~\cite{BerkeleyABC} incorporated in RDF does not support complex sequencing elements.
This library also comes with technology and cell LEF files, which can be readily used for all the placement and routing stages.

\subsection{Circuit Netlists}
A set of circuit netlists are taken from the TAU 2017 Timing Contest~\cite{tau2017} as well as from the IWLS 2005 Benchmarks~\cite{iwls2005}.
They are remapped to the standard cell libraries described in the previous section, and the most critical path delay of each circuit is measured.
To set tight timing constraints, the clock period is set to $80\%$ of the critical path delay for each circuit.
The number of cells in the netlists range from 352 to 571853.

\subsection{Detailed Placement}
The first EDA research contest, ISPD 2005 contest, focused on mixed-size cell placement~\cite{Nam:2005gx}.
In the past decade, most of research endeavors have been devoted to global placement, and current state-of-the-art placers become mature.
Very recently, detailed placement and legalization request novel ideas to handle mixed-cell-height circuits for better power, area, routability, and performance trade-offs.
A legalizer removes all cell overlaps, meets complicated design rules and constraints, and preserves the ``good'' solution provided by global placement as much as possible.
Considering the mixed-cell-height standard cell designs with various design rules at advanced technology nodes, 2017 ICCAD held a Mixed-Cell-Height Standard Cell Legalization Contest~\cite{Darav:2017}.

In RDF, the recent mixed-cell-height legalizer~\cite{Zhu:2018} which won the first place award of the contest is included.

\begin{table}[!t]
    \caption{Design Rules and Routing Preference Metrics.}
    \vspace{-2mm}
    \small
    \begin{tabularx}{0.925\columnwidth}{@{}YY@{}}
    \toprule
    Design rules                & Routing preference metrics \\ \midrule
    Open                        & Wrong-way routing          \\
    Short                       & Off-track routing          \\
    Parallel run length spacing & Routing guide honoring     \\
    End of line spacing         &                            \\
    Cut spacing                 &                            \\
    Min area rule (MAR)         &                            \\ \bottomrule
    \end{tabularx}
    \label{tab:design_rules}
\end{table}

\begin{figure*}[!t]%
    \centering
    \includegraphics[width=\textwidth]{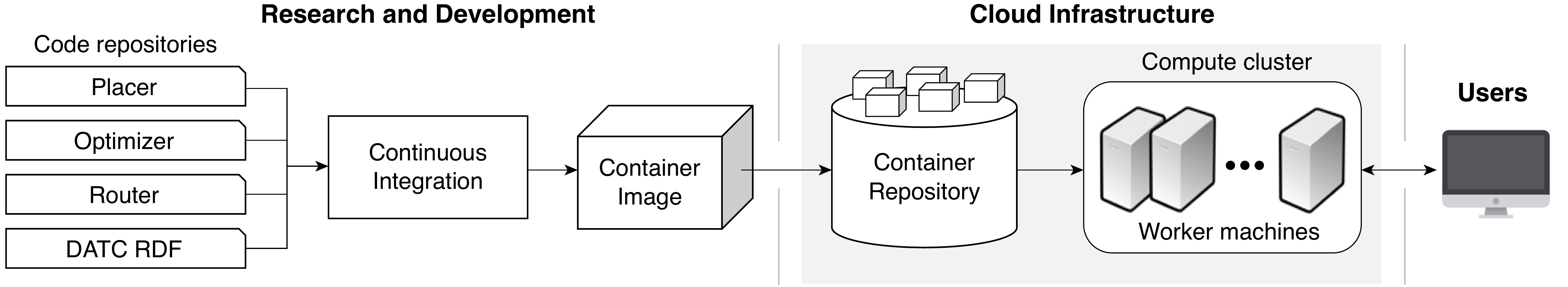}%
    \caption{A conceptual illustration of DATC RDF in scalable cloud infrastructure. Source codes of individual point tools and DATC RDF are maintained in source code repositories, which are continuously integrated and containerized using a continuous integration framework. The resulting container image is then deployed into cloud infrastructure, which can be accessed by end-users.}%
    \label{fig:rdf_cloud}%
\end{figure*}

\subsection{Detailed Routing}
ISPD 2018 Initial Detailed Routing Contest~\cite{Mantik:2018is} is the first contest that targets detailed routing considering practical design rules and honoring global routing guidance.
DATC RDF is extended to accommodate the outcome of the detailed routing contest.

In RDF, global routing and detailed routing read input files based on ISPD 2008 Global Routing Contest~\cite{Nam:2008gr} and ISPD 2018 Initial Detailed Routing Contest~\cite{Mantik:2018is}, respectively.
Since there is no industrial standard format for connecting global routing and detailed routing, we develop a global routing guide translator to translate the output format of ISPD 2008 Global Routing Contest into the input format of routing guide used in ISPD 2018 Initial Detailed Routing Contest.
In ISPD 2018 Contest, a group of design rules and routing preference metrics are defined (\Cref{tab:design_rules}) and stored in LEF/DEF files.
As in commercial routers, the output of a detailed router follows DEF format that can be read by any commercial layout tools.

Currently, NCTUdr is included, and more tools from winning teams will be included.

%------------------------------------------------------------------------------
% DATC RDF Cloud
%------------------------------------------------------------------------------
\section{DATC RDF in Scalable Cloud Infrastructure}\label{sec:datc_rdf_cloud}
Because of the today's crises of design complexity, quality, and cost, a truly new approach and paradigm of design tools and flows are highly required~\cite{ABK:2018op, OpenROAD}.
To foster such research efforts to the open-source cloud-based CAD tools based on previous CAD contests, we propose a development flow and an implementation of RDF~\cite{datc_rdf_cloud}, which can be readily deployed especially in the scalable cloud infrastructure.
It consists of three fundamental parts as illustrated in~\Cref{fig:rdf_cloud}: code repositories, continuous integration and containerization, and container orchestration.
We expect that they make it easier to collaborate the development of point tools and the design flow, and to deploy the entire system in someone's own machine or public cloud infrastructure.

Source codes of each point optimization tool are maintained in source code repositories, such as git and Mercurial; RDF itself is also maintained in the code repositories.
They are then integrated and \textit{containerized} into a container image~\cite{Merkel:2014do}, which can be automatically done by continuous integration tools~\cite{Meyer2014:co}.
As the source codes are maintained using source code repositories and continuous integration tools, the implementation of RDF can stay always up to date.

With the containerization, one can readily deploy the entire DATC RDF framework because the container keeps all the necessary libraries and dependencies that are necessary to run RDF.
In particular, current mainstream cloud providers, such as \textit{Amazon AWS}~\cite{AWS}, \textit{Microsoft Azure}~\cite{Azure}, and \textit{IBM Cloud}~\cite{IBMCloud} IBM Cloud, all provide off-the-shelf solutions for automated container deployment engine, e.g., Kubernetes.
Besides, scaling the deployment can be easily achieved with the ready-to-use horizontal scaling and load balancing features of container orchestration systems.
We also expect that the containerization will realize large-scale parallel architectures of design automation systems.

%------------------------------------------------------------------------------
% Experimental results
%------------------------------------------------------------------------------
 \section{Experiments and Demonstration}\label{sec:exp}
DATC RDF framework is implemented using C++ and Python3.
We demonstrate our flow based on a benchmark circuit \texttt{fft\_ispd} from the TAU 2017 Timing Contest~\cite{tau2017}.

\begin{figure}[!t]%
        \centering
        \includegraphics[width=0.95\linewidth]{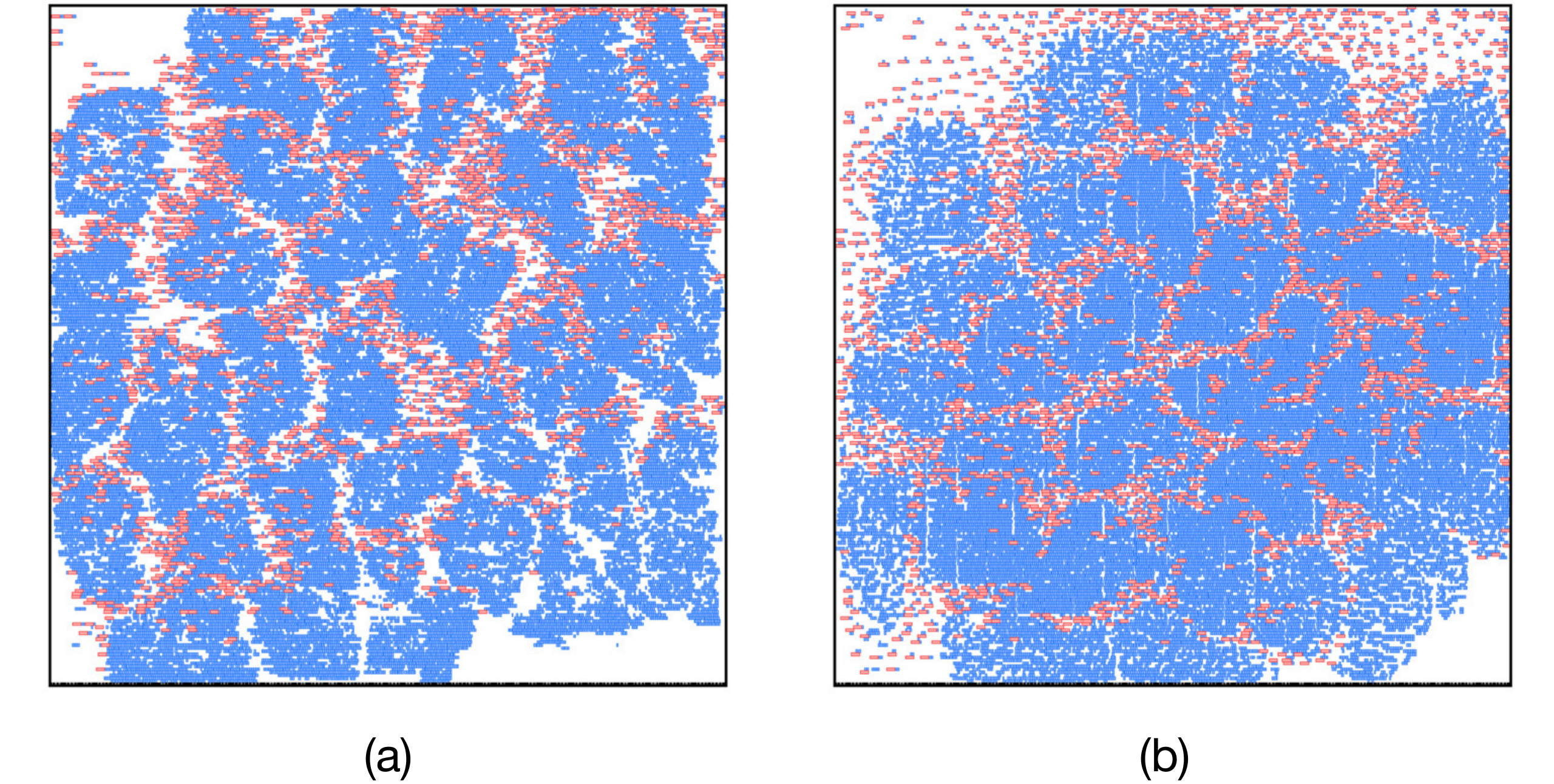}%
    \caption{Placement results of fft\_ispd after logic synthesis with the Lazy-man script of ABC. The sequencing elements are represented as the red boxes, and the combinational gates are as the blue boxes. (a) ComPLx and (b) NTUPlace3.}%
        \label{fig:placement_plot}%
    \end{figure}

\begin{figure}[!t]%
        \centering
        \includegraphics[width=0.95\linewidth]{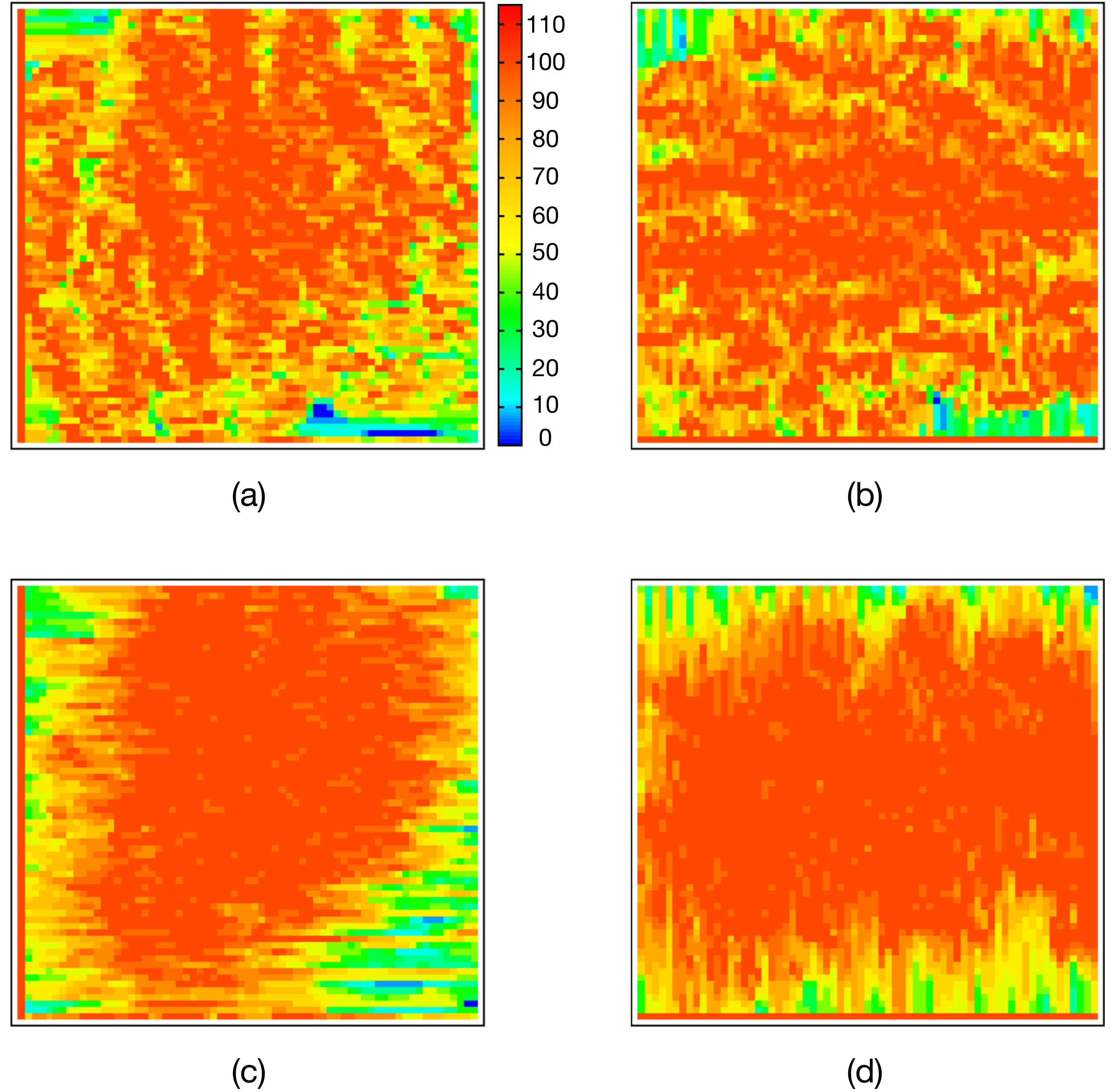}%
        \caption{Global routing congestion map of fft\_ispd. Placement result is obtained using ComPLx placer, and global routing is done by NCTU-GR 2.0. (a) Metal-3, (b) Metal-4, (c) Metal-5 and (d) Metal-6 layers.}
        \label{fig:congestion_map}%
    \end{figure}

\begin{figure}[!t]%
        \centering
        \includegraphics[width=0.95\linewidth]{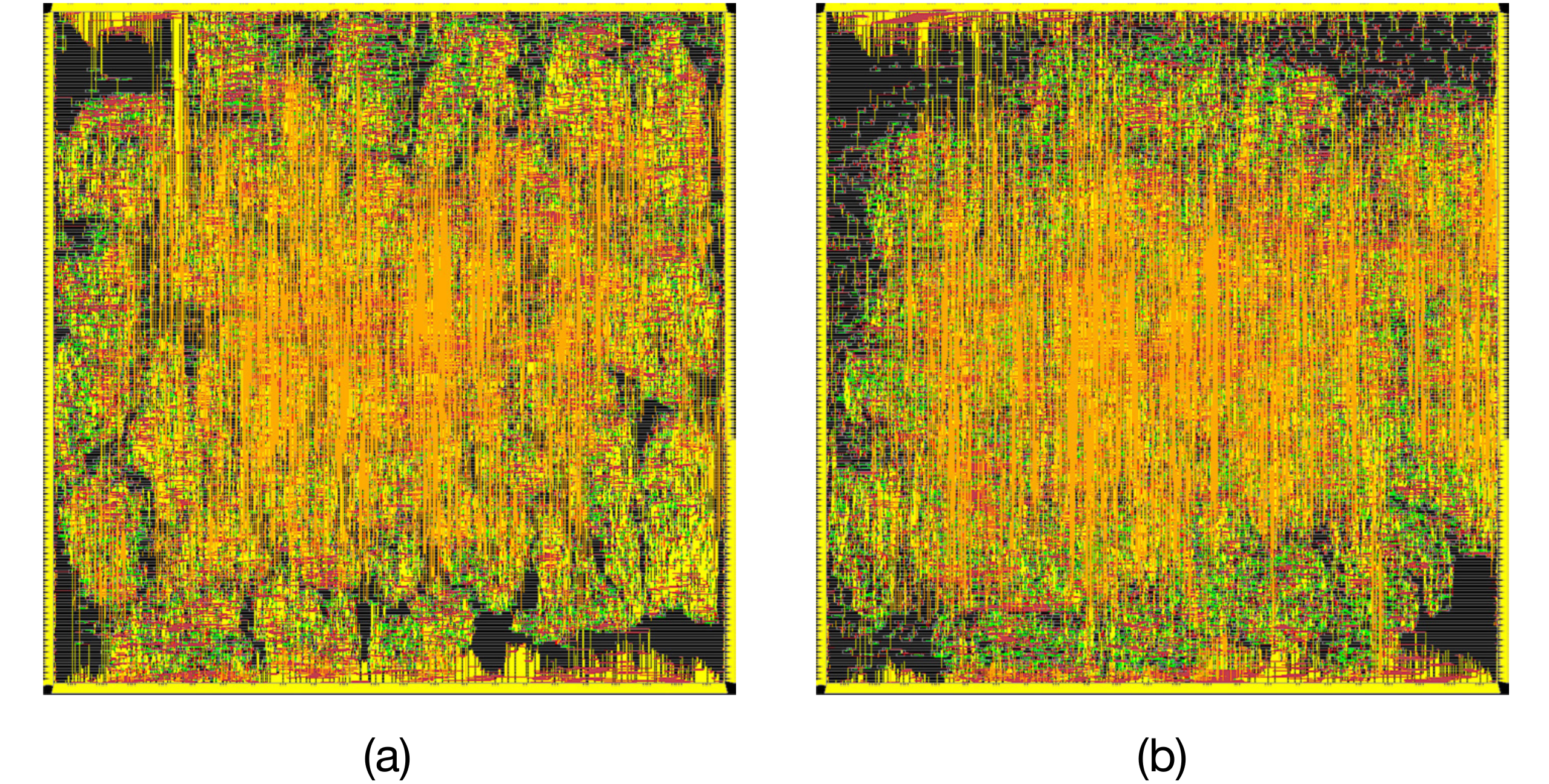}%
        \caption{Detailed routing results of fft\_ispd. Metal-3 to Metal-6 layers are colored with green, yellow, red, and orange, respectively. Placement is done with (a) ComPLx and (b) NTUPlace3.}%
    \label{fig:detailed_route}%
    \end{figure}

The circuit netlist was first unmapped to a generic gate library, and subsequently remapped to our standard cell library using \textit{Synopsys DesignCompiler L-2016.03-SP5-5}~\cite{DC}.
It was then synthesized using the ABC logic synthesis and verification platform~\cite{BerkeleyABC} using the AIG optimization script of Lazy-man synthesis paradigm~\cite{Yang:2012lm}.
Two placement instances were then created using ComPLx~\cite{Kim:2012jx} and NTUPlace3~\cite{Chen:2008fd}.
They were then routed with NCTU-GR 2.0~\cite{Liu:2013hp} and BFG-R~\cite{Hu:2010ct} global routers.
Finally, detailed routing was performed with NCTUdr.
The results are shown in~\Cref{fig:placement_plot}, \Cref{fig:congestion_map}, and \Cref{fig:detailed_route}.

\section{Conclusion}\label{sec:conclusion}
In this paper, we present DATC RDF, which is an open design flow from logic synthesis to detailed routing. We include point tools based on previous EDA research contests and will keep expanding the flow coverage vertically and horizontally. We also demonstrate RDF in cloud infrastructure. RDF can be readily integrated with design methodology and cross-stage optimization research.

\section*{Acknowledgment}
This work was supported by IEEE CEDA Design Automation Technical Committee (DATC). Special thanks go to tool providers for their generosity.

\bibliographystyle{ACM-Reference-Format}
\bibliography{references}

\end{document}